\documentclass[conference]{IEEEtran}
\IEEEoverridecommandlockouts
\usepackage{svg}
\usepackage{comment}
\usepackage{cite}
\usepackage{amsmath,amssymb,amsfonts}
\usepackage{algorithmic}
\usepackage{graphicx}
\usepackage{textcomp}
\usepackage{xcolor}
\usepackage[a4paper, total={184mm,239mm}]{geometry}
\def\BibTeX{{\rm B\kern-.05em{\sc i\kern-.025em b}\kern-.08em
    T\kern-.1667em\lower.7ex\hbox{E}\kern-.125emX}}
    
\def\omitauthors #1{{\textit{Authors’ name omitted for double-blind review}}}
\def\omitacks #1{{\textit{Acknowledgement omitted for double-blind review}}}

\begin{document}

\title{End-to-End DNN Inference on a Massively Parallel Analog In Memory Computing Architecture}

\author{\IEEEauthorblockN{Nazareno Bruschi\IEEEauthorrefmark{1}, Giuseppe Tagliavini\IEEEauthorrefmark{1}, Angelo Garofalo\IEEEauthorrefmark{1}\IEEEauthorrefmark{2}, Francesco Conti\IEEEauthorrefmark{1},}\IEEEauthorblockN{Irem Boybat\IEEEauthorrefmark{3}, Luca Benini\IEEEauthorrefmark{1}\IEEEauthorrefmark{2}, Davide Rossi\IEEEauthorrefmark{1}}
\IEEEauthorblockA{
\IEEEauthorrefmark{1}\textit{University of Bologna}, Bologna, Italy, \IEEEauthorrefmark{2}\textit{ETH}, Zurich, Switzerland}
\IEEEauthorrefmark{3}\textit{IBM Research}, Zurich, Switzerland}

\maketitle

\begin{abstract}

The demand for computation resources and energy efficiency of Convolutional Neural Networks (CNN) applications requires a new paradigm to overcome the “Memory Wall”. Analog In-Memory Computing (AIMC) is a promising paradigm since it performs matrix-vector multiplications, the critical kernel of many ML applications, in-place in the analog domain within memory arrays structured as crossbars of memory cells. However, several factors limit the full exploitation of this technology, including the physical fabrication of the crossbar devices, which constrain the memory capacity of a single array. Multi-AIMC architectures have been proposed to overcome this limitation, but they have been demonstrated only for tiny and custom CNNs or performing some layers off-chip. In this work, we present the full inference of an end-to-end ResNet-18 DNN on a 512-cluster heterogeneous architecture coupling a mix of AIMC cores and digital RISC-V cores, achieving up to 20.2 TOPS. Moreover, we analyze the mapping of the network on the available non-volatile cells, compare it with state-of-the-art models, and derive guidelines for next-generation many-core architectures based on AIMC devices.
\end{abstract}


\section{Introduction}\label{sec:introduction}

Matrix-Vector Multiplication (MVM) is the critical operation in modern Deep Neural Networks (DNN), and its optimization has been tackled from different perspectives, from software kernels to hardware accelerators.
In recent years, Analog In-Memory Computing (AIMC) has been a widely studied computing paradigm since it promises outstanding performance and energy efficiency on MVM operations~\cite{sebastian2020memory}.
However, the large-scale usage of AIMC in commercial products is limited by technological issues, especially in fabricating large arrays. AIMC can be employed using very different memory technologies, which can be classified as volatile and non-volatile.

The former has a more mature community, especially for SRAM technology, due to its robustness and viability for large-scale integration in any CMOS node. Several SRAM-based chips have been developed targeting any DNN requirements~\cite{seo2022digitalvsanalog}. However, they generally require moving network parameters among large off-chip memories to be temporarily stored in the on-chip computational memory, negatively impacting energy consumption~\cite{hung2021challenges}.

Non-volatile AIMC (nvAIMC) instead merges parameter storage with computational memory. In this way, parameters do not need to be transferred from on- or off-chip storage through the memory hierarchy. 
However, the limited writing access speed~\cite{yu2021compute-in-memory} of nvIMC devices introduces the need for a static mapping strategy to preserve the performance capability of such devices.
Moreover, a further challenge is the fabricable size of nvIMC devices, which de-facto is limited to 1024$\times$1024 with up to 8-bit equivalent memory cells~\cite{yu2021compute-in-memory}.

During the last few years, the fabrication of several prototypes exploiting these technologies~\cite{jia2021programmable, papistas2021a22nm, khaddam2022hermes} demonstrated the feasibility of the approach, despite several open challenges related to the intrinsic variability of analog computing, the need for specialized training to address analog noise and non-idealities. On the other hand, most of these works aimed at demonstrating the technology rather than targeting end-to-end inference of deep neural networks. One of the limitations of nvAIMC cores is their little flexibility due to the ability to implement only MVMs.

For this reason, few recent works~\cite{jia2020heterogeneousmicroprocessor, garofalo2022heterogenous, zhou2021analognets} proposed the integration of nvAIMC cores into digital System-on-Chips (SoC), exploiting a mix of nvAIMC cores and more flexible specialized and programmable digital processors. Thanks to this mix, they demonstrated remarkable performance on the full inference of neural networks in the mobile domain, such as MobileNetV2, time multiplexing computations on several nvAIMC cores~\cite{garofalo2022heterogenous, zhou2021analognets}. Indeed, in the mobile domain, it is common that only one sample is processed at a time, relaxing the requirements of layer pipelining. This constraint significantly limits the potential of nvAIMC since only one core can be active at a given time.

This constraint can be relaxed when leaving the mobile domain: high-performance inference of DNNs typically exploits batching due to the large number of images typically processed in HPC and data centers applications. Several recent works exploited this feature proposing many-core data-flow architectures. On the other hand, most of these works made strong assumptions about the characteristics of the networks to be processed to better fit the shape of the DNN on the proposed architectures. For example, Dazzi et al.~\cite{dazzi2021efficientpipelined} targeted a relatively small ResNet-like network targeting the CIFAR-10 dataset, while Shafiee et al.~\cite{shafiee2016isaac} and Ankit et al.~\cite{ankit2019puma} target VGG-like networks featuring no residual layers, nicely fitting mapping on pipelined data-flow architectures.
However, DNNs generally feature data flow graph loops (e.g., residual layers) that make a straightforward pipelining implementation much more challenging. Moreover, most of these architectures only feature specialized accelerators for implementing digital functions such as ReLU, and MaxPool, somehow limiting the flexibility of their approach.

In this work, we tackle the problem from another perspective. We present a general-purpose system based on RISC-V cores for digital computations and nvAIMC cores for analog-amenable operations, such as 2D convolutions. A scalable hierarchical network-on-chip interconnects the system to maximize on-chip bandwidth and reduce communication latency.
We evaluate all the system inefficiencies, especially for the non-ideal mappings and communication infrastructure bottleneck running a real-life network for state-of-the-art applications such as ResNet-18 inference on 256$\times$256 image dataset. We perform an experimental assessment on an extended version of an open-source system-level simulator~\cite{bruschi2021gvsoc}, resulting in up to 20.2 TOPS and 6.5 TOPS/W for the whole ResNet-18 inference of a batch of 16 256x256 images in 4.8 ms. The hardware and software described in this work are open-source, intending to support and boost an innovation ecosystem for next-generation computing platforms.

\section{Massively Parallel Heterogeneous System Architecture}\label{sec:architecture}

\begin{figure}[t]
\centering
\includegraphics[scale=0.34]{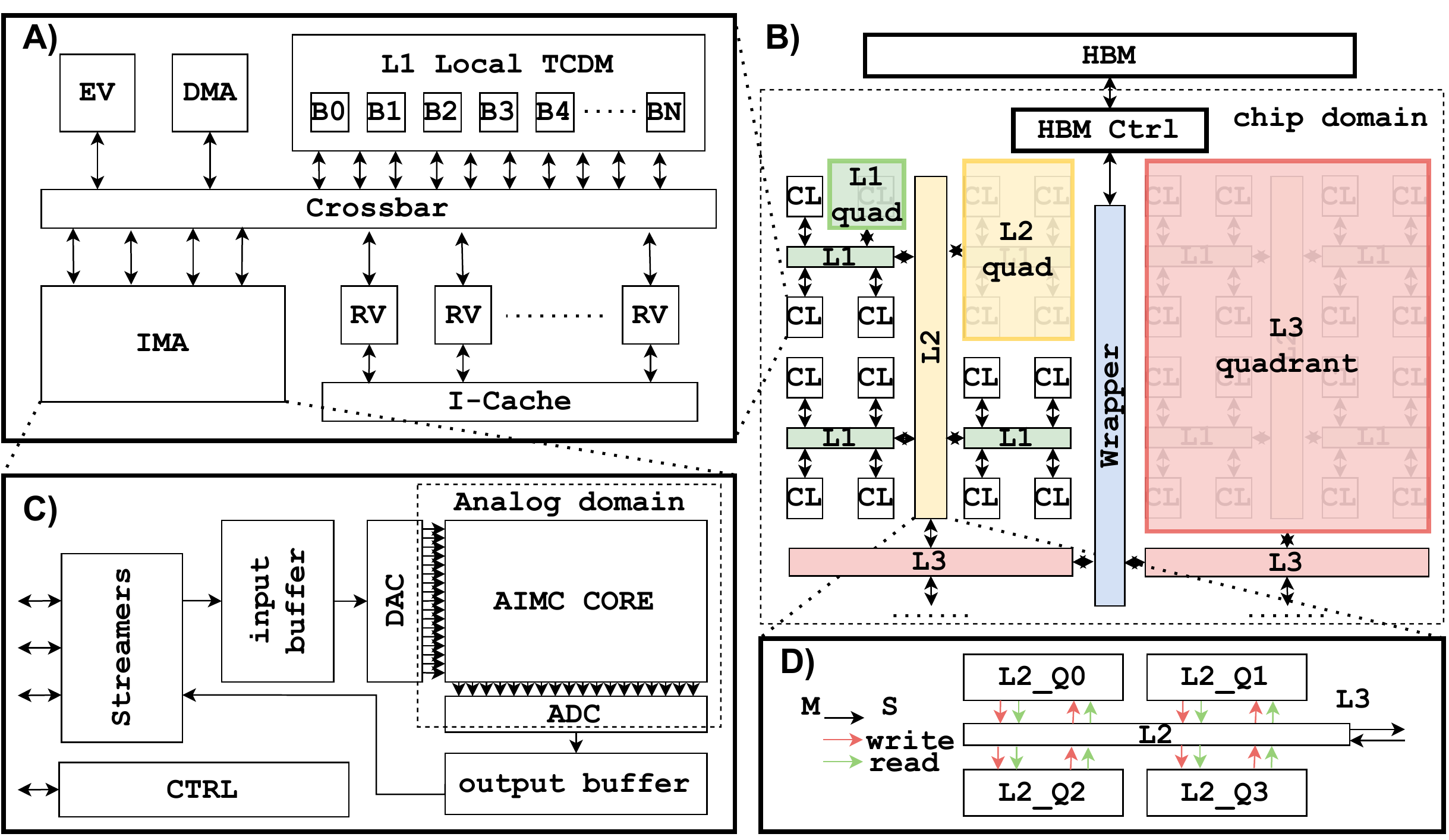}
\caption{A) Cluster architecture. B) Massively parallel system architecture. C) IMA subsystem. D) Router model.
}
\label{fig:architecture}
\end{figure}

This section presents the proposed heterogeneous many-core SoC architecture. It consists of multiple heterogeneous (analog/digital) clusters communicating through a hierarchical AXI interconnect gathering data from a shared High-Bandwith Memory (HBM), as shown in Fig.~\ref{fig:architecture}B.

\subsubsection{\textbf{Cluster}}\label{subsec:cluster} The core of the proposed system architecture consists of heterogeneous analog/digital clusters (Fig.~\ref{fig:architecture}A). Each cluster contains a set of RISC-V cores (CORES)~\cite{gautschi2017pulp}, a shared multi-bank scratchpad data memory (L1) enabling Single Program Multiple Data (SPMD) computations, a hardware synchronizer to accelerate common parallel programming primitives such as thread dispatching and barriers, and a DMA for the cluster to cluster and cluster to HBM communication. Each cluster also includes a nvAIMC Accelerator (IMA) sharing the same multi-banked memory as the CORES for efficient communication, similarly to the architecture presented in Garofalo et al.~\cite{garofalo2022heterogenous}.

\subsubsection{\textbf{IMA}}\label{subsec:imc accelerator} The IMA is built around a Phase-Change Memory (PCM) computational memory organised as a 2D array featuring horizontal \textit{word lines} and vertical \textit{bit lines} (Fig.~\ref{fig:architecture}C). In computational memory, the PCM cells are exploited as programmable resistors placed at the cross points between the word lines and the bit lines, which allows the implementation of MVM in the analog domain with high parallelism and efficiency. In this work, we assume an MVM to be executed in 130~ns as reported in Khaddam et al.~\cite{khaddam2022hermes}. At the beginning of each \textit{word lines} and the end of each \textit{bit lines}, Digital-to-Analog (DAC) and Analog-to-Digital converters (ADC) converters perform the conversion between analog and digital domains, respectively. ADCs and DACs connect to two digital buffers connected to the L1 memory through a set of streamers featuring programmable address generation.

\subsubsection{\textbf{Interconnect}}\label{subsec:interconnect} The interconnect infrastructure connecting the clusters in the proposed many-core architecture consists of a highly parametrizable \textit{hierarchical network} composed of a set of AXI4 \textit{nodes}, as proposed in Kurth et al.~\cite{kurth2021anopensourceplatform}. The network topology specifies different regions called quadrants connecting groups of clusters: the \textit{Level 1} nodes connect $N_{1}$ quadrants (clusters), the \textit{Level 2} nodes connect $N_{2}$ \textit{Level 1} quadrants, and the Level \textit{Level N} nodes connect $N_{N}$ \textit{Level N-1} quadrants, as shown in Fig.~\ref{fig:architecture}B. The \textit{Quadrant Factor} for a given \textit{level N} defines the number of quadrants (either clusters or level N-1 quadrants) connected to the AXI node for each level. Clusters feature a master and a slave port, which means that a transaction can either be initiated by the target cluster through its master port or by any other cluster through the target cluster’s slave port. The same concept applies to the whole hierarchy of quadrants. In both cases, transactions can be either read or write transactions with full support for bursts according to AXI4 specifications. The last level of the interconnect architecture, called Wrapper, connects all the levels below to the off-chip HBM through an HBM controller.

\section{Simulation Infrastructure}\label{sec:simulation infrastructure}

\begin{table}[t]
\caption{GVSoC architecture parameters}
\begin{center}
\begin{tabular}{|c|c|}
\hline
\textbf{Parameter} & \textbf{Value} \\
\hline
Number of clusters & 512 \\
\hline
Number of IMA per cluster & 1 \\
\hline
Number of CORES per cluster & 16 \\
\hline
L1 memory size & 1 MB \\
\hline
HBM size & 1.5 GB \\
\hline
Operating frequency & 1 GHz \\
\hline
Number of streamers ports (read and write) & 16 \\
\hline
IMA crossbar size & 256$\times$256 \\
\hline
Analog latency (MVM operation) & 130 ns \\
\hline
Quadrant factor (HBM link,wrapper,L3,L2,L1) & (1,8,4,4,4) \\
\hline
Data Width (HBM link,wrapper,L3,L2,L1) & (64,64,64,64,64) Bytes \\
\hline
Latency (HBM, link,wrapper,L3,L2,L1) & (100,4,4,4,4) cycles \\
\hline
\end{tabular}
\label{tab:params}
\end{center}
\end{table}


We modeled the proposed architecture by extending an open-source simulator named GVSOC~\cite{bruschi2021gvsoc} meant to simulate RISC-V-based clustered multi-core architectures. It is a C++ event-based simulator featuring a simulation speed of 25 MIPS and an accuracy of more than 90\% compared to a cycle-accurate equivalent architecture when simulating a full DNN in a single cluster, as reported in Bruschi et al.~\cite{bruschi2021gvsoc}. 

The main components integrated into the simulator are the IMA and the interconnect infrastructure extending the capabilities of the simulator towards many-core accelerators (i.e., up to 512 clusters and 8192 RISC-V cores). The IMA is integrated into the cluster as a master of the cluster crossbar. All the components of the IMA have been modeled, including the input and output buffers and the streamers. At the system level, the interconnect infrastructure has been modeled as a set of parametric router components with configurable data width, latency, and the number of master and slave ports combined together to create the topology described in Fig.~\ref{fig:architecture}D. Table~\ref{tab:params} describes the configuration parameters of the platform used in this work. All the modules in the simulator have been calibrated using the cycle-accurate RTL and FPGA equivalent. 256$\times$256 IMA size has been used since it has been demonstrated in more works and shows better technological feasibility at this time~\cite{khaddam2022hermes}. This infrastructure allows to simulate the execution of a full ResNet-18 on 512 instantiated clusters in less than 20 minutes on a 32 GB RAM, Intel(R) Core(TM) i7-2600 CPU @ 3.40GHz.

\section{Computational model}\label{sec:computational model}

\begin{figure}[t]
\centering
\includegraphics[scale=0.3]{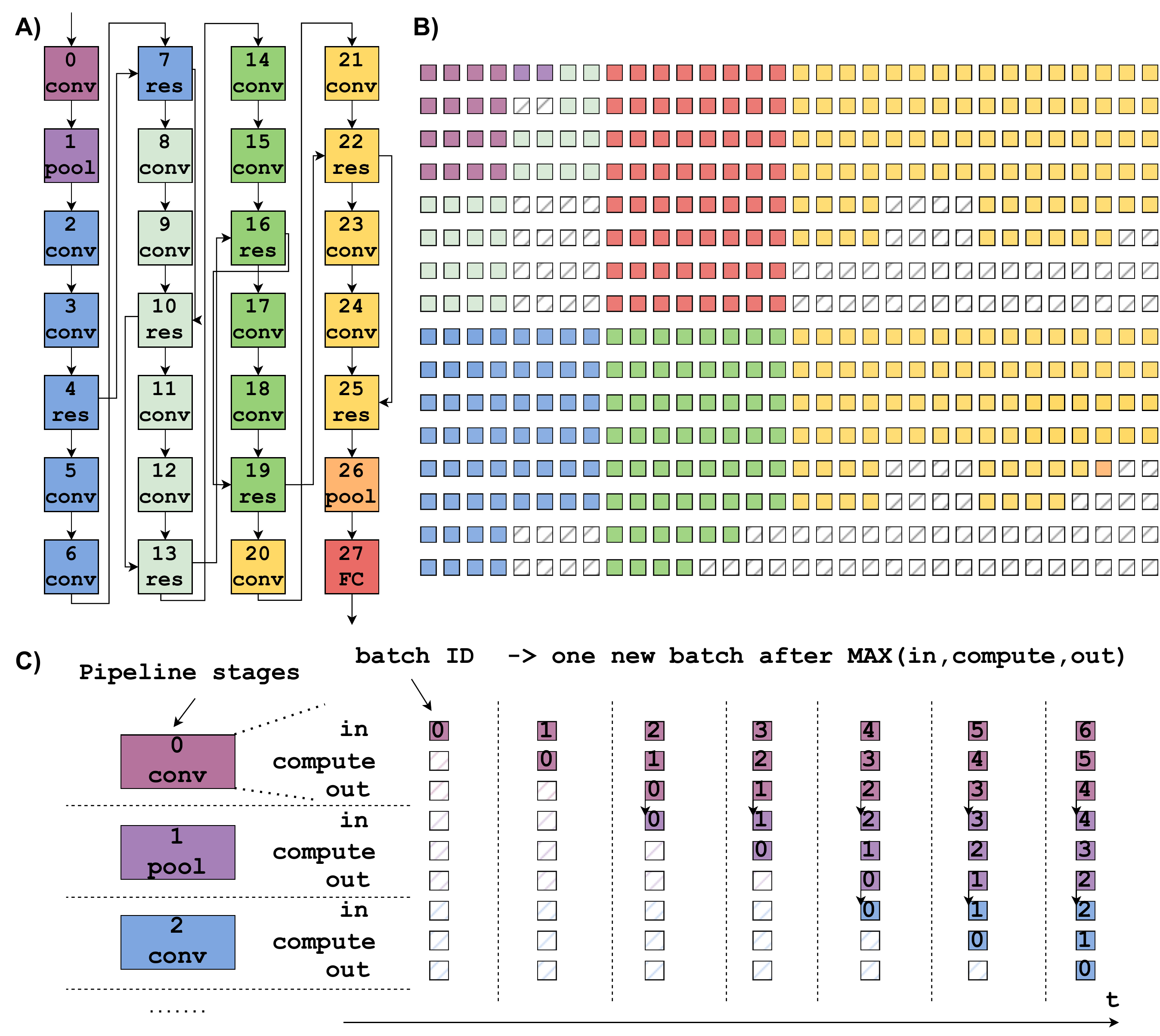}
\caption{A) Directed Acyclic Graph (DAG) of the ResNet-18 execution. B) Mapping example on 512 clusters. C) High-level description of pipelining computational model.}
\label{fig:dag}
\end{figure}

This section presents the computational model of the proposed massively parallel heterogeneous architecture, detailing its main characteristics: Layer Mapping, IMA execution, Pipelining, Data Tiling, and Self-Timed Execution Flow.

\subsubsection{\textbf{Static Layer Mapping}}\label{subsec:mapping}

According to the computational model of the proposed many-core architecture, each layer of a DNN is statically mapped to a certain number of clusters, while the input/output features maps (IFM/OFM) are streamed from producer to consumer clusters. Fig.~\ref{fig:dag}B shows the mapping of the ResNet-18 on the architecture, where each node of the graph in Fig.~\ref{fig:dag}A represents a CNN layer, grouped by color according to the IFM dimensions, and every layer is mapped on different clusters of the system, as shown in Fig.~\ref{fig:dag}B. The number of clusters used to map a specific layer depends on the number of parameters of the layer. For example, \textit{Layer 22} features 2.3M parameters, requiring 40 clusters for the mapping, assuming each 256x256 IMA can store 64K parameters.

\subsubsection{\textbf{IMA Execution}}\label{subsec:ima}

\begin{figure}[t]
\centering
\includegraphics[scale=0.33]{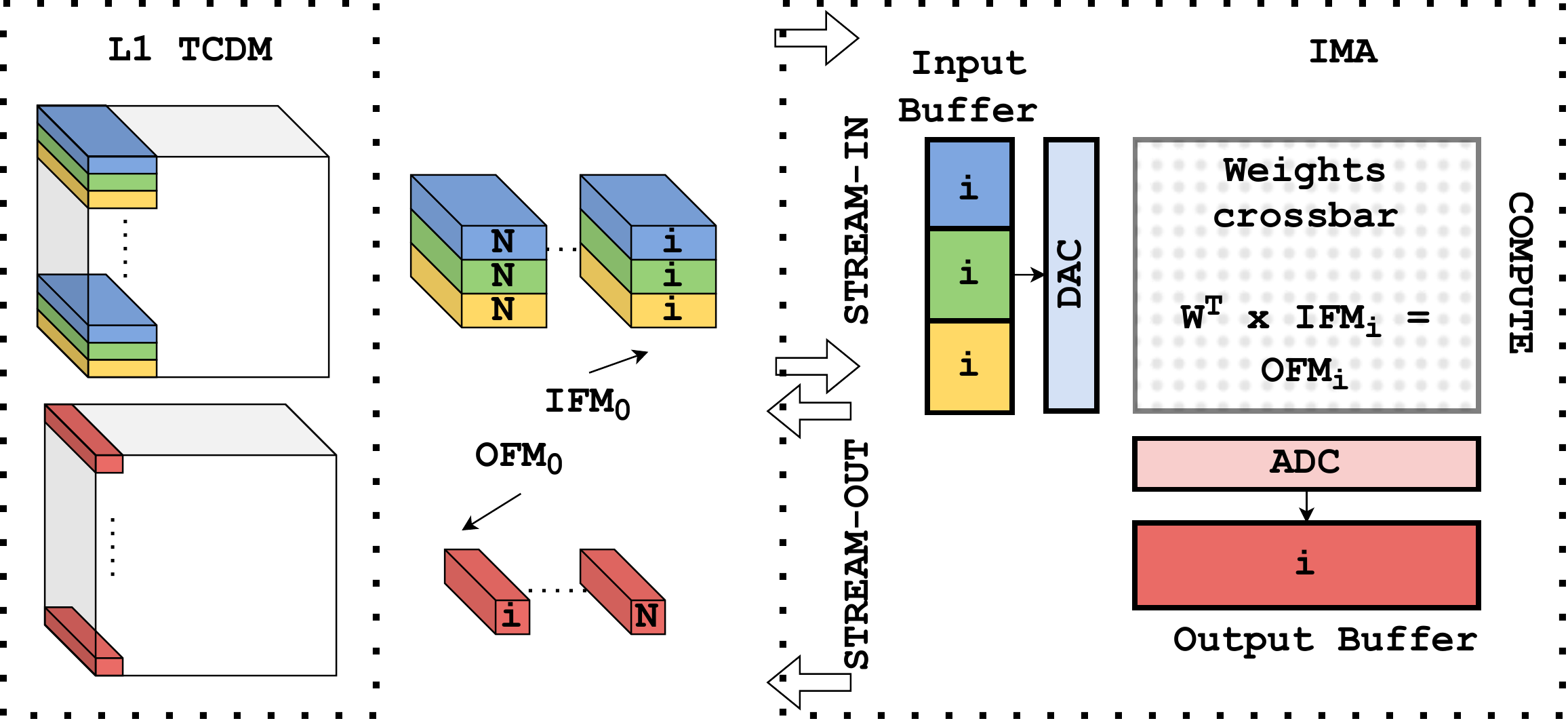}
\caption{IMA execution model.}
\label{fig:ima}
\end{figure}

As described in Sec.~\ref{sec:architecture}, the IMA subsystem communicates directly to the L1 of the cluster, acting as a master of the TCDM interconnect. Assuming DNN parameters of a specific layer are being pre-loaded to the non-volatile array, IMA execution is composed of three distinct phases, as shown in Fig.~\ref{fig:ima}. \textit{Stream-in} fetches the IFM of the layer and moves them to the input buffer of the IMA. \textit{Compute} performs the input data conversion by the DACs, the analog MVM execution on the crossbar, and the ADCs conversion. \textit{Stream-out} moves the output digital MVM result from the output buffers to the L1 memory. Input and output buffers are duplicated to enable double buffering, completely overlapping the cost of transfers between the L1 and the buffers with the computation, maximizing the computational efficiency of the accelerator.

\subsubsection{\textbf{Pipelining}}\label{subsec:pipelining}

When the inference starts, the IFM of the first layer is streamed into the first set of clusters which process it generating the OFM, which is then passed to the second set of clusters and so on. Assuming the possibility of having large batches of images allows for the creation of the software pipeline described in Fig.~\ref{fig:dag}C, where different chunks of data are processed by a different set of clusters simultaneously, fully overlapping the data movements (i.e., in charge of the DMA) with the computation (i.e., in charge of IMA and/or CORES). Ensuring that all pipeline stages execute in the same amount of time is essential when creating such a pipeline structure. Techniques to speed-up pipeline stages, such as \textit{parallelization} and \textit{data-replication}, will be discussed in Sec.~\ref{sec:evaluation}.

\subsubsection{\textbf{Data Tiling}}\label{subsec:tiling}

To fit IFM/OFM of large DNN models within the limited memory resources of the clusters (1~MB of L1 memory is assumed in this work), we split IFM/OFM into smaller chunks of data called \textit{tiles}, processed by the clusters as soon as the input data is transferred to the L1 memory. In particular, data tiling is always performed along the $W_{in}$ and $W_{out}$ dimensions for input and output, respectively. In this work, we assume a static tiling strategy, and $W_{in/out}$ implicitly defines the batching dimension. Therefore, the batches are composed of vertical slices of IFM/OFM. The other dimensions ($C_{in}$ and $H_{in}$) are, when necessary, tiled in other clusters to fit the memory requirements (parameters mapping) or to speed up the computation (\textit{parallelization}).

\subsubsection{\textbf{Self-Timed Execution}}\label{subsec:self-timed}

To implement the pipeline between the tiled structure described in \ref{subsec:tiling}, we exploit a data-flow self-timed execution model. Computation in a cluster can be performed by the CORES, IMA, or both in parallel. While software execution on the CORES is synchronous, IMA execution is managed asynchronously (like DMA transfers). A cluster can perform a certain computation whenever three conditions are satisfied: a) Chunk N+1 from the \textit{producers} can be loaded to the L1 memory, b) the \textit{consumers} are ready to accept the output data of chunk N-1, c) both IMA and CORES are free to compute chunk N. If all the conditions are satisfied, the new iteration can start with the following execution flow: 1) the CORE0 (i.e., master core) first waits for the events from the input and output DMA channels and IMA, 2) the CORE0 configures and triggers I/O DMA channels and IMA for computation of next tile 3) digital processing is performed in parallel on the CORES. 4) All the CORES go to sleep, waiting for the events described in point 1).

\section{Evaluation: ResNet-18}\label{sec:evaluation}

\begin{figure}[t]
\centering
\includegraphics[scale=0.4]{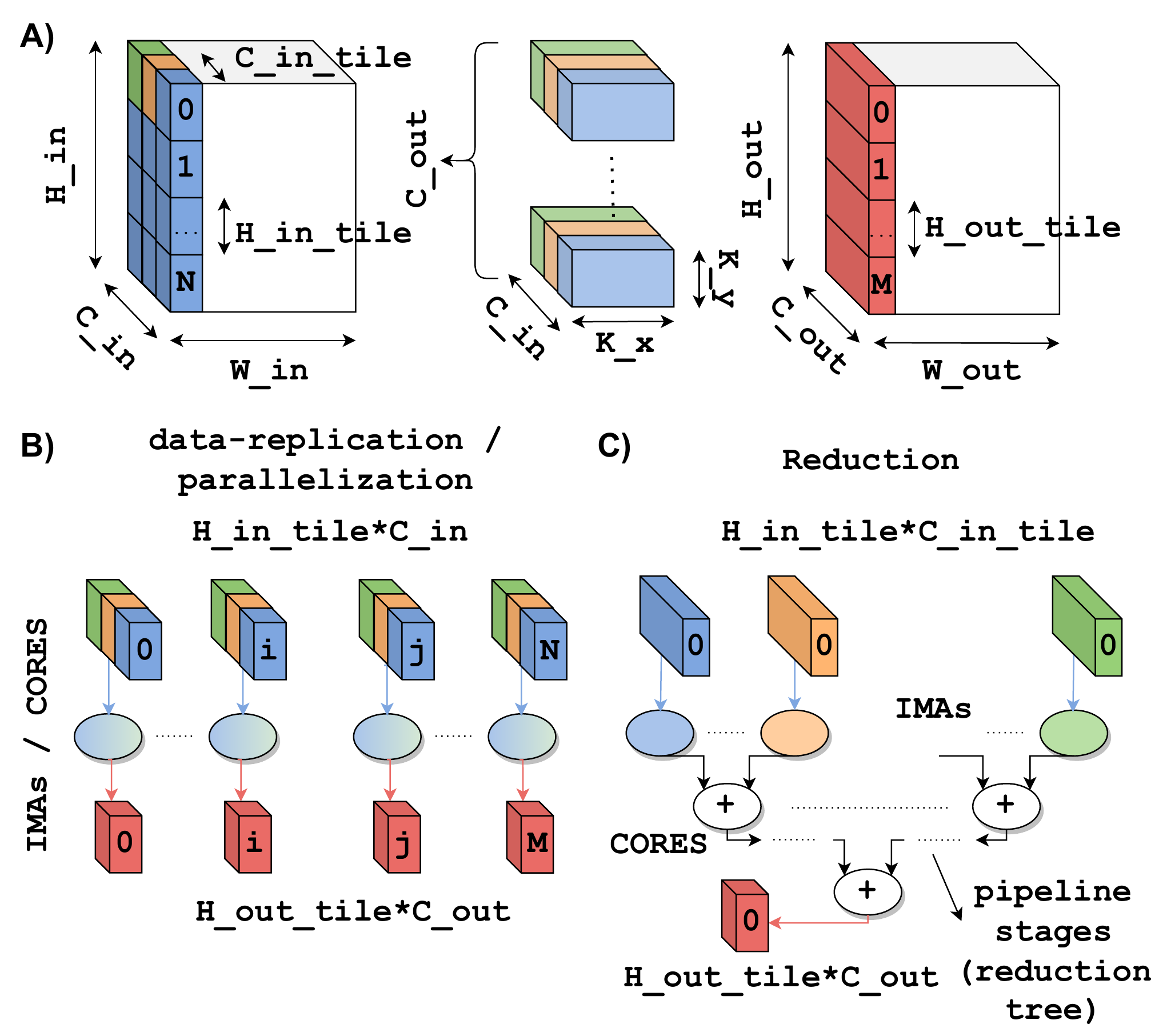}
\caption{A) Generic layer IFM, parameters and OFM. B) \textit{Data-replication} and \textit{parallelization}. C) Reduction operation.}
\label{fig:mapping implementation}
\end{figure}

In this section, we present the mapping of the ResNet-18 on the proposed many-core architecture, providing insights on the adaptations required to the baseline mapping and execution model presented in Sec.~\ref{sec:computational model} to map all the layers and balance the pipeline optimally. The key operation of a ResNet-18 is a sequence of two 3x3 2D convolutions followed by a tensor addition (i.e., residual layer) between the OFM of the previous layer and the OFM of the previous residual layer. The first two layers are a 7x7 2D convolution followed by a MaxPool activation layer that starts propagating the residuals. The topology is shown in Fig.~\ref{fig:dag}A.

\subsubsection{\textbf{Multi-cluster layers}}\label{subsec:multi-cluster layers}

When mapping a real-life network to the many-core architecture, the ideal condition would consist of perfectly matching the parameters of every layer with the size of the IMA. Unfortunately, this is not the case in the most general case. Two types of situations might arise, depending on the dimension of the IFM/OFM. When the size of the input channels multiplied by the kernel size ($C_{in}\times K_{x}\times K_{y}$) is greater than the number of rows of the IMA (i.e., 256), multiple IMAs are needed to compute the partial outputs. Then, a reduction among these partial outputs has to be performed to compute the OFM. On the other hand, while the size of the output channels is larger than the number of columns of the IMA (i.e., 256), the inputs need to be broadcasted to all the IMA involved (storing a different set of output channels parameters), and multiple IMAs compute part of the output channels at the same time. In some layers of ResNet-18, the two situations arise concurrently. This mapping approach has to be applied to all the layers computed in the analog domain, excluding \textit{Layer 0}.

\subsubsection{\textbf{Data-replication and Parallelization}}\label{subsec:data-replication}

In a pipelined computational model, the throughput of the whole pipeline is limited by the latency of the slowest stage. Hence, the pipeline has to be balanced as much as possible to achieve high throughput. Unbalancing might depend on many causes, and we will explain some of them in Sec.~\ref{sec:results}. A technique to speed up the execution of layers (i.e., one stage of the pipeline) executed in the analog domain (i.e., on the IMA) is \textit{data-replication}. \textit{Data-replication} increases the parallelism, replicating the parameters of a layer on different IMAs and computing at the same time multiple jobs on multiple chunks of the IFM. With this approach, the speed-up, net of overheads due to communication and data tiling, is theoretically equal to the number of replications at the cost of area, as the same layer parameters are stored on multiple IMAs (Fig.~\ref{fig:mapping implementation}B). In this work, we extensively use this technique, especially for the first layers of the network. 
If the bottleneck of the pipeline is a layer executed in the digital domain (e.g., residual, reductions, pooling), one option is to parallelize the computation on the CORES over multiple clusters. A plain parallelization scheme 
is used for pooling and residual layers (i.e., Layers 1, 4, 7, 13, 19).

\begin{figure}[t]
\centering
\includegraphics[scale=0.64]{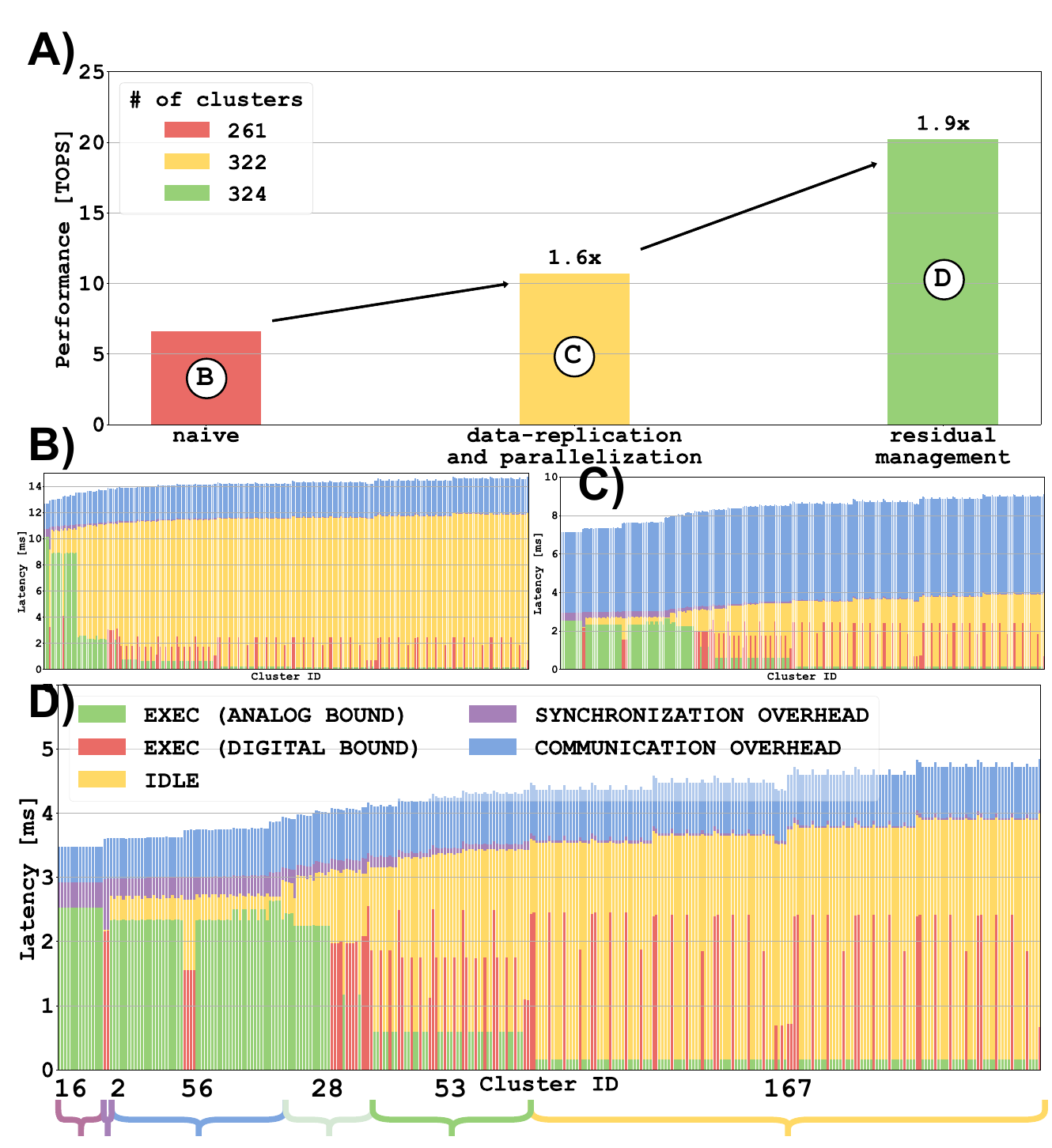}
\caption{ResNet-18 inference results. A) Throughput with different mapping optimizations. B) Execution time on every cluster in naive implementation. C) Execution time on every cluster in \textit{data-replication} and \textit{parallelization} implementation. D) Execution time on every cluster in final implementation.}
\label{fig:results}
\end{figure}

\subsubsection{\textbf{Reduction Management}}\label{subsec:reduction}

A different approach has to be adopted for the reduction since this operation requires a hierarchical tree (Fig.~\ref{fig:mapping implementation}C) featuring limited and decreasing parallelism. In particular, the level of parallelism of this operation is implicit in the structure of the network. In ResNet-18, we might have to sum up the partial products of up to 20 clusters (i.e., Layers 20-21, 23-24) according to the multi-cluster mapping strategy described in Sec.~\ref{subsec:multi-cluster layers}. In this context, the computation of the residuals might form a bottleneck for the pipeline since this operation has to be performed by the CORES in the digital domain. To accelerate these layers we split the hierarchical tree into several pipeline stages and assign each pipeline stage to a logarithmically decreasing number of clusters with well-balanced latency. This approach has been exploited in all reduction layers.

\subsubsection{\textbf{Residuals Management}}\label{subsec:residual}

In an ideal pipelined data flow, data are exchanged only among consecutive pipeline stages. On the other hand, in many modern DNNs such as ResNet-18, this is not the case due to the presence of residual layers. Unfortunately, with limited resources in terms of memory storage (1~MB per cluster), and considering the residual's data lifetime between when it is produced and when it is consumed, external temporary storage has to be used to store this temporary data. In particular, in our pipeline, ResNet-18 requires 1.6 MB to simultaneously store all the residuals of the whole network, where the minimum dimension can be calculated as $C_{out}*H_{out}$. While a first intuitive approach to tackle this issue is to exploit the off-chip HBM memory due to its large capacity, exploiting such memory as temporary storage for residual blocks significantly increases the traffic towards this high-latency memory controller, forming a bottleneck for the whole pipeline reducing its overall performance. Instead, a better solution is to exploit the L1 memory of clusters not used for computations for residual storage, improving the performance compared to the baseline approach by 1.9$\times$.

\section{Results and Discussion}\label{sec:results}

\begin{figure}[t]
\centering
\includegraphics[scale=0.11]{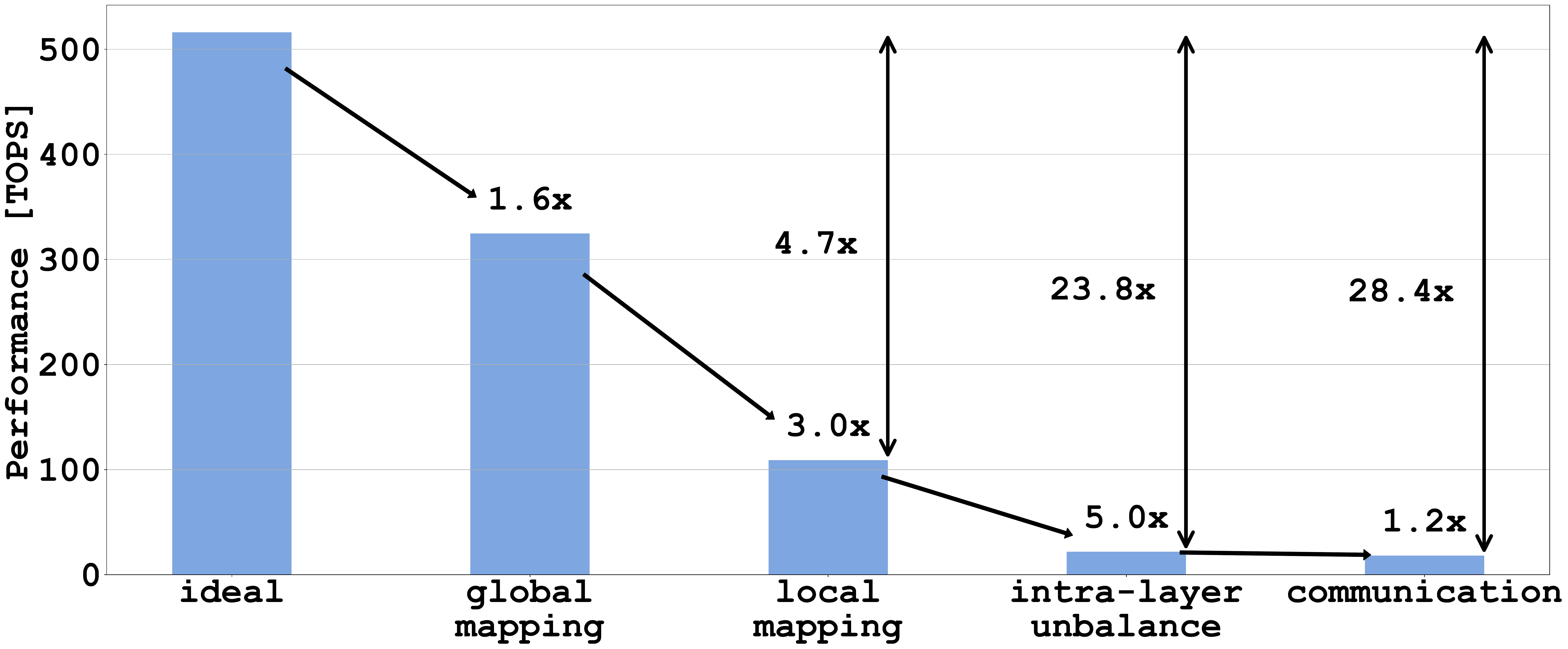}
\caption{Performance degradation considering non-idealities due to static mapping, network topology, and communication.}
\label{fig:consideration}
\end{figure}


In this section, we analyze the results of ResNet-18 execution mapped on the proposed many-core heterogeneous architecture. To extract reliable physical implementation information from the architecture, we performed the physical implementation (down to ready for silicon layout) of the cluster in 22nm FDX technology from Global Foundries. We used Synopsys Design Compiler for physical synthesis, Cadence Innovus for Place\&Route, Siemens Questasim for extracting the value change dump for activity annotation, and Synopsys PrimeTime for power analysis. Area, frequency, and power figures are then scaled to a 5nm tech node more suitable for modern HPC architectures. Fig.~\ref{fig:results}D shows the execution time of a batch of 16 256$\times$256 images on the architecture. For each cluster, it shows the amount of time spent on computation, communication, synchronization, and sleeping. Since analog and digital computations are performed in parallel, execution bars are indicated in green when analog bound and in red when digital bound. We can note an expected increasing trend of latency with the cluster ID caused by the head and tail of the pipeline execution (i.e., idle times waiting for the first and the last batches are propagated through the pipeline).

Fig.~\ref{fig:results}A shows the performance gain achieved thanks to the techniques described in Sec.~\ref{sec:computational model}. Fig.~\ref{fig:results}B shows the latency breakdown of the clusters in a naive implementation, where all the network parameters are mapped into the architecture exploiting the multi-cluster technique described (Sec.~\ref{subsec:multi-cluster layers}) but with no further optimizations. It is possible to note the large unbalance between the first layers and the deeper layers in the network. Fig.~\ref{fig:results}C shows the latency breakdown after \textit{data-replication} and \textit{parallelization}, better balancing the pipeline and improving performance by 1.6$\times$ at the cost of utilising 61 more clusters. Reducing compute latency moves the bottleneck of the execution on communication due to large contentions on the HBM, mainly caused by residual management. Fig.~\ref{fig:results}d shows the optimized mapping of residual described in Sec.~\ref{subsec:residual} further improving performance by 1.9$\times$ at the cost of 2 more clusters (to exploit 2 MB of available on-chip memory). 

\begin{figure}[t]
\centering
\includegraphics[scale=0.11]{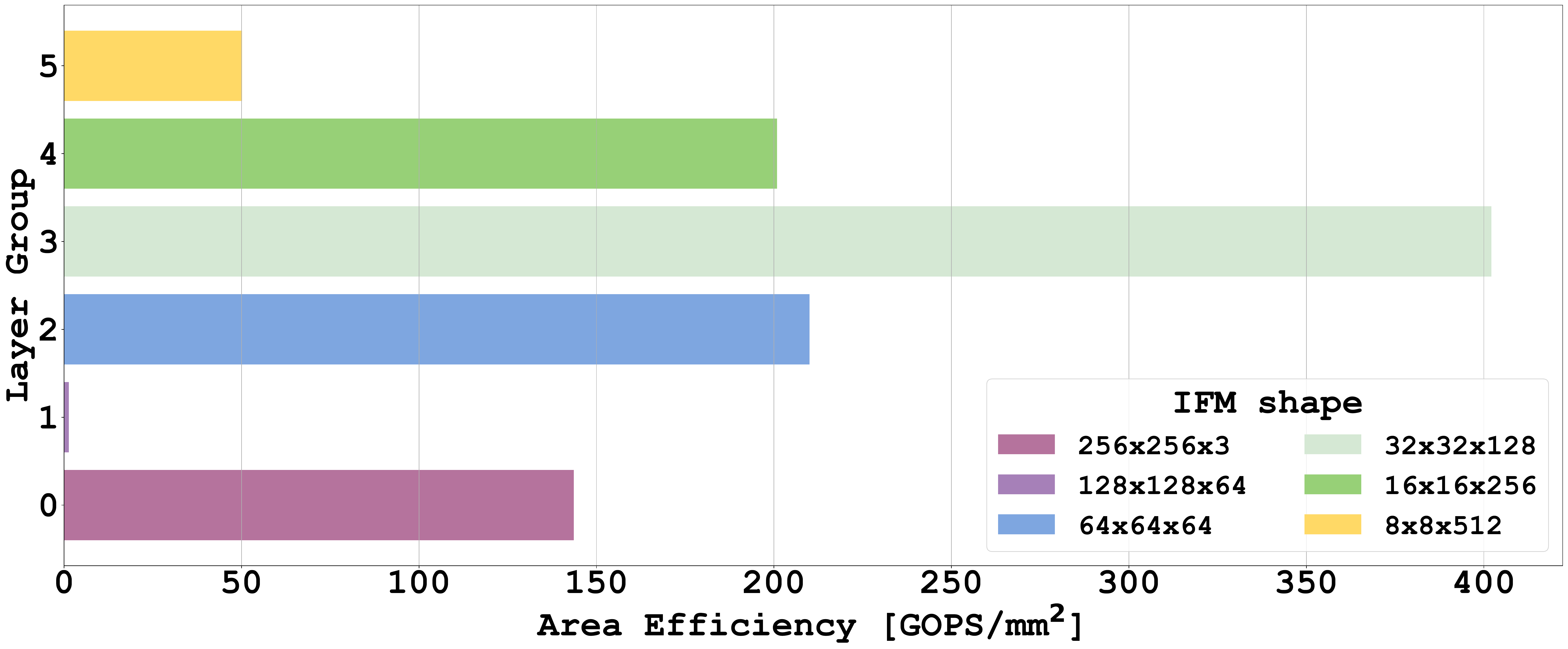}
\caption{Area efficiency per group of clusters as defined in Fig.~\ref{fig:dag} without communication inefficiencies.}
\label{fig:exploration}
\end{figure}

To provide insights into the sources of inefficiency highlighted in Fig.~\ref{fig:results}D, we analyze the mapping and latency breakdown of the Resnet-18 inference. The first source of inefficiency (global mapping) is caused by the fact that not all the clusters are used for mapping network parameters. In our mapping, 322 clusters out of 512 have been exploited. This is an intrinsic characteristic of all systolic architectures exploiting pipelining as a computational model, worsened by the constraints in terms of mapping imposed by IMA. However, this has only an effect on the area efficiency since, in such regular architecture, each cluster can be easily clock and power gated, minimizing the impact on energy efficiency. The second source of inefficiency (local mapping) is caused by the fact that even if a specific cluster is being used, the mapping on it might under-utilize the analog and digital resources. In some cases, parameters cannot fill the whole IMA; in other cases, the array is not used at all. The same happens for digital computing, e.g., in the case of purely digital layers. A possible solution to mitigate this degradation could be to integrate heterogeneous clusters configured to fit better all the possibilities, such as IMA and a single CORE (i.e., analog clusters) or 16 CORES without IMA (i.e., digital clusters).

The third source of inefficiency is caused by the pipeline unbalance. Different layers feature different computational efficiency, as described in Fig.~\ref{fig:exploration}, where the layer groups are defined depending on the IFM dimension. Some layer groups feature significant area efficiency, thanks to large IFM/OFM implying high data reuse (i.e., several iterations over the same parameters statically mapped on the IMA). In particular, Layer 12 (i.e., group 3) is executed on 10 clusters, with \textit{data-replication} factor of 2, leading to a peak of efficiency of 600~GOPS/mm${}^2$. Conversely, deeper layers in the network, because of the stride, feature analog layers with very poor parameters reuse interleaved with stages of reductions executed by the CORES. In particular, Layer 20, 21, 23, and 24 (i.e., group 5) are executed on 40 clusters each. This causes extremely low latency for the execution of the analog layers (less than 0.2~ms), which translates into lower area efficiency (50~GOPS/mm${}^2$) compared to the first layers. A possible approach to tackle this inefficiency might lead to further exploiting heterogeneity by coupling IMA and CORES with a set of more compact specialized digital accelerators more suitable for low-data reuse layers, improving the silicon efficiency. Another approach could be to use larger IMA arrays~\cite{Narayanan2021fullyon-chip}. However, this would require more data transfers per cluster.

Despite the analyzed sources of inefficiency, the proposed architecture delivers 20.2~TOPS (i.e., 3303 ~images/s) and 42~GOPS/mm${}^2$ on the end-to-end inference of Resnet-18. Performing the inference in 9.2~ms and 15~mJ, which corresponds to an energy efficiency of 6.5 TOPS/W, paves the way for a new generation of general-purpose many-core architectures exploiting a mix of analog and digital computing.

\section{Conclusion}\label{sec:conclusions}

In this work, we have proposed a general-purpose heterogeneous multi-core architecture based on a PULP cluster augmented with nvAIMC accelerators to efficiently execute real end-to-end networks, exploiting the throughput of such a paradigm. We have proposed a mapping based on the combination of pipelining execution flow and many techniques to increase the parallelism and split the workload among the nvAIMC cores. We have shown the results of the inference of a batch of 16 256$\times$256 images on a ResNet-18, obtaining up to 20.2 TOPS and 6.5 TOPS/W for a 480 mm2 architecture. We have finally provided an exhaustive performance analysis, considering a real case of traffic and highlighting the criticisms when nvAIMC is used in real applications, providing several insights on how to mitigate this effect, to drive the design and the usage of nvAIMC architecture as a general-purpose platform for DNN acceleration.

\section*{Acknowledgement}

This work was supported by the WiPLASH project (g.a. 863337), founded by the European Union’s Horizon 2020 research and innovation program.

\bibliography{bibliography}
\bibliographystyle{ieeetr}

\end{document}